\documentclass{emulateapj}
\usepackage{apjfonts}
\usepackage{graphicx}
\usepackage{xspace}

\newcommand{\Msun}      {\mbox{$\rm\,M_{\mathord\odot}$}}

\newcommand\gso{{GSO}\xspace}
\newcommand\hxd{{HXD}\xspace}
\newcommand\pin{{PIN}\xspace}
\newcommand\xis{{XIS}\xspace}

\newcommand\cyg{{Cyg~X-1}\xspace}

\begin{document}

\lefthead{{\em NuSTAR} and {\em Suzaku} Observations of Cyg X-1}
\righthead{Tomsick et al.}

\submitted{Accepted by ApJ}

\def\lsim{\mathrel{\lower .85ex\hbox{\rlap{$\sim$}\raise
.95ex\hbox{$<$} }}}
\def\gsim{\mathrel{\lower .80ex\hbox{\rlap{$\sim$}\raise
.90ex\hbox{$>$} }}}

\title{The reflection component from Cygnus X-1 in the soft state measured by
{\em NuSTAR} and {\em Suzaku}}

\author{John A. Tomsick\altaffilmark{1}, Michael A. Nowak\altaffilmark{2}, Michael Parker\altaffilmark{3}, Jon M. Miller\altaffilmark{4}, Andy C. Fabian\altaffilmark{3}, Fiona A. Harrison\altaffilmark{5}, Matteo Bachetti\altaffilmark{6,7}, Didier Barret\altaffilmark{6,7}, Steven E. Boggs\altaffilmark{1}, Finn E. Christensen\altaffilmark{8}, William W. Craig\altaffilmark{1,9}, Karl Forster\altaffilmark{5}, Felix F\"{u}rst\altaffilmark{5}, Brian W. Grefenstette\altaffilmark{5}, Charles J. Hailey\altaffilmark{10}, Ashley L. King\altaffilmark{4}, Kristin K. Madsen\altaffilmark{5}, Lorenzo Natalucci\altaffilmark{11}, Katja Pottschmidt\altaffilmark{12,13}, Randy R. Ross\altaffilmark{14}, Daniel Stern\altaffilmark{15}, Dominic J. Walton\altaffilmark{5}, J\"{o}rn Wilms\altaffilmark{16}, and William W. Zhang\altaffilmark{17}}

\altaffiltext{1}{Space Sciences Laboratory, 7 Gauss Way, University of California, Berkeley, CA 94720-7450, USA (e-mail: jtomsick@ssl.berkeley.edu)}

\altaffiltext{2}{Massachusetts Institute of Technology, Kavli Institute for Astrophysics, Cambridge, MA 02139, USA}

\altaffiltext{3}{Institute of Astronomy, University of Cambridge, Madingley Road, Cambridge CB3 0HA, UK}

\altaffiltext{4}{Department of Astronomy, University of Michigan, 500 Church Street, Ann Arbor, MI 48109, USA}

\altaffiltext{5}{Cahill Center for Astronomy and Astrophysics, California Institute of Technology, Pasadena, CA 91125, USA}

\altaffiltext{6}{Universit\'{e} de Toulouse; UPS-OMP; IRAP; Toulouse, France}

\altaffiltext{7}{CNRS; Institut de Recherche en Astrophysique et Plan\'{e}tologie; 9 Av. colonel Roche, BP 44346, F-31028 Toulouse cedex 4, France}

\altaffiltext{8}{DTU Space, National Space Institute, Technical University of Denmark, Elektrovej 327, DK-2800 Lyngby, Denmark}

\altaffiltext{9}{Lawrence Livermore National Laboratory, Livermore, CA 94550, USA}

\altaffiltext{10}{Columbia Astrophysics Laboratory, Columbia University, New York, NY 10027, USA}

\altaffiltext{11}{Istituto Nazionale di Astrofisica, INAF-IAPS, via del Fosso del Cavaliere, 00133 Roma, Italy}

\altaffiltext{12}{CRESST and NASA Goddard Space Flight Center, Astrophysics Science Division, Code 661, Greenbelt, MD 20771, USA}

\altaffiltext{13}{Center for Space Science and Technology, University of Maryland Baltimore County, 1000 Hilltop Circle, Baltimore, MD 21250, USA}

\altaffiltext{14}{Physics Department, College of the Holy Cross, Worcester, MA 01610, USA}

\altaffiltext{15}{Jet Propulsion Laboratory, California Institute of Technology, Pasadena, CA 91109, USA}

\altaffiltext{16}{Dr.~Karl-Remeis-Sternwarte and Erlangen Center for Astroparticle Physics, Sternwartestrassa 7, D-96049, Bamberg, Germany}

\altaffiltext{17}{NASA Goddard Space Flight Center, Greenbelt, MD 20771}

\begin{abstract}

The black hole binary Cygnus~X-1 was observed in late-2012 with the 
{\em Nuclear Spectroscopic Telescope Array (NuSTAR)} and {\em Suzaku},
providing spectral coverage over the $\sim$1--300\,keV range.  The
source was in the soft state with a multi-temperature blackbody, 
power-law, and reflection components along with absorption from 
highly ionized material in the system.  The high throughput of 
{\em NuSTAR} allows for a very high quality measurement of the complex 
iron line region as well as the rest of the reflection component.  
The iron line is clearly broadened and is well-described by a 
relativistic blurring model, providing an opportunity to constrain 
the black hole spin.  Although the spin constraint depends somewhat 
on which continuum model is used, we obtain $a_{*}$$>$0.83 for all models 
that provide a good description of the spectrum.  However, none of 
our spectral fits give a disk inclination that is consistent with 
the most recently reported binary values for Cyg~X-1.  This may 
indicate that there is a $>$13 degree misalignment between the orbital 
plane and the inner accretion disk (i.e., a warped accretion disk) or 
that there is missing physics in the spectral models.

\end{abstract}


\section{Introduction}

Cygnus~X-1 is a bright high-mass X-ray binary that was discovered in
the early days of X-ray astronomy \citep{bowyer65} and was identified
with the optical counterpart HD~226868 \citep{mw71}.  It is best known 
for being the first system with a high enough mass measurement to rule 
out the possibility that the compact object is a neutron star 
\citep[e.g.,][]{gb86}, making it the first confirmed black hole (BH) 
system.  The current constraint on the BH mass is 
$14.8\pm 1.0$\Msun~\citep{orosz11}.  Cyg~X-1 has been instrumental in 
improving our understanding of accreting BHs, their spectral states, 
and the relationship between the accretion disk and the jet 
(see Remillard \& McClintock 2006\nocite{rm06} for a review of BH binaries).

Currently, a major on-going effort in BH studies is to measure their spins.
A non-zero BH spin changes the space-time around the black hole, requiring 
the Kerr rather than the Schwarzchild metric to describe the geometry.  
The spin is also one possible source for powering the relativistic jets 
seen coming from BHs.  One technique for measuring the BH spin involves 
modeling the multi-temperature thermal component that comes from the 
accretion disk \citep{mcclintock06}.  A major challenge for this technique 
is that the distance to the system and the inclination of the inner disk 
must be known.  For Cyg~X-1, the distance is well-established with a 
parallax measurement of $1.86^{+0.12}_{-0.11}$\,kpc \citep{reid11}, 
which is consistent with a measurement using the dust scattering halo 
\citep{xiang11}.  The improved distance determination has also led to new 
constraints on the binary inclination.  Combined modeling of optical 
spectroscopy (i.e., the companion's radial velocity) and photometry over 
all orbital phases has given a binary inclination of $27.1\pm 0.8$~degrees 
\citep{orosz11}.  Orbital modulations are seen in the optical light curves
that depend on the shape of the companion star and the inclination of the
system.  Although some misalignment between the inner disk inclination and 
the binary inclination is possible \citep{maccarone02}, under the assumption 
that they are the same, \cite{gou11} find that the spin of the Cyg~X-1 BH 
is $a_{*}$$>$0.92 (3-$\sigma$ limit), and an even higher spin limit 
($a_{*}$$>$0.983 at 3-$\sigma$) has been recently reported \citep{gou13}.

Another technique for measuring BH spin involves modeling the Compton 
reflection component that is due to hard X-ray emission shining on the 
inner part of the optically thick accretion disk.  The reflection spectrum
includes fluorescent emission lines, with the Fe K$\alpha$ lines typically 
being the strongest \citep{fabian89}, and a broad excess in the $\sim$10--50\,keV 
energy range \citep{lw88}.  The reflection spectrum can be distorted by the 
relativistic effects of Doppler broadening from the fast orbital motion and 
the gravitational redshift due to the BH's gravitational field \citep{fabian89}.  
The emission lines can also be broadened when photons are Compton-scattered 
out of the narrow line core \citep{rf05}.  This implies that broad emission 
lines are not necessarily an indication of relativistic effects.  However, 
the Compton-broadening is symmetric, so modeling the asymmetric component is 
the key to using this technique to constrain BH spin \citep{rn03,miller07}.

For both thermal and reflection component modeling techniques, the BH spin
measurement is actually inferred from the measurement of the location of 
the inner radius of the optically thick and ``cold'' (i.e., not fully ionized)
accretion disk.  The BH spin measurement then comes from identifying the inner 
radius with the innermost stable circular orbit (ISCO).  For a non-rotating BH 
($a_{*}\equiv Jc/GM_{\rm BH} = 0$, where $J$ is the angular momentum of the BH, 
$c$ is the speed of light, $G$ is the gravitational constant, and $M_{\rm BH}$ is 
the mass of the BH), the ISCO is at 6 gravitational radii 
($R_{\rm g} = GM_{\rm BH}/c^{2}$), and, for a maximally rotating BH ($a_{*} = 1$), 
the ISCO approaches 1\,$R_{\rm g}$.

For Cyg~X-1, most of the reflection studies have used X-ray spectra 
from times when the source was in the hard state.  In this state, 
it is unclear whether the assumption about the inner disk radius being 
at the ISCO holds.  For BH transients, studies allow for the possibility
that the disk recedes when the source is in the faint hard state at an 
Eddington fraction ($L/L_{\rm Edd}$) of $\sim$0.1--0.01\% \citep{nwd02,tomsick09c,cabanac09}, 
but there is evidence that the disk remains close to or at the ISCO during 
the bright part of the hard state \citep{miller06a,rfm10}.  Historically, 
Cyg~X-1 has been in the bright part of the hard state, making the ISCO 
assumption plausible.  Using hard state observations, \cite{nowak11} 
did not report a spin measurement but put an upper limit on disk 
recession.  Other reflection-based measurements constrained the BH 
spin to be $0.6\leq a_{*} \leq 0.99$ \citep{miller12}, $a_{*} = 0.88^{+0.07}_{-0.11}$ 
\citep{duro11}, and $a_{*} = 0.97^{+0.014}_{-0.02}$ \citep{fabian12}.

The reflection fits in the hard state provide evidence for high BH spin 
consistent with the limit on the BH spin from thermal modeling in the
soft state.  In this paper, we report on the details 
of reflection modeling in the soft state using observations with the 
{\em Nuclear Spectroscopic Telescope Array} \citep{harrison13}
and {\em Suzaku} \citep{mitsuda07}.  {\em NuSTAR} covers the 3--79\,keV 
bandpass, which is ideal for reflection studies.  Its detectors 
give unprecedented energy resolution in the hard X-ray band, and
provide high throughput without the photon pile-up that occurs for
charge-coupled device (CCD) observations of bright sources.  {\em NuSTAR} 
has already been used for reflection studies of the supermassive BH 
NGC~1365 \citep{risaliti13} as well as the Galactic BH GRS~1915+105
\citep{miller13}.  In this paper, we provide details of the 
observations, instrument capabilities, and the data reduction 
methods in \S\,2.  The results of the spectral fitting are reported 
in \S\,3, and the results are discussed in \S\,4.  Finally, we 
present conclusions in \S\,5.

\section{Observations and Data Reduction}

\begin{figure}
\includegraphics[scale=0.47]{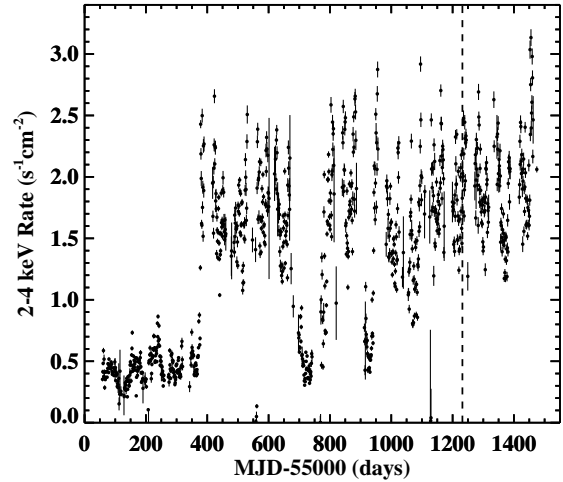}
\caption{\small {\em MAXI} light curve in the 2--4\,keV band for Cyg~X-1 
between mid-2009 and mid-2013.  The source was in the hard state until 
MJD 55,377 and has spent most of its time in the soft state since then.  
The {\em NuSTAR} and {\em Suzaku} observation that is the subject of 
this work is indicated with a vertical dashed line.\label{fig:lc_maxi}}
\end{figure}

We observed Cyg~X-1 with {\em NuSTAR} and {\em Suzaku} on 2012 October 31 
and November 1 (MJD 56,231 and 56,232).  Figure~\ref{fig:lc_maxi} shows the
soft X-ray light curve from the {\em Monitor of All-sky X-ray Image}
\citep[{\em MAXI};][]{matsuoka09}, indicating how this observation fits into the 
$\sim$4\,yr history of this source.  At the time of the observation, the {\em MAXI} 
2--4\,keV count rate (normalized by effective area) was 
$1.83\pm 0.04$\,s$^{-1}$\,cm$^{-2}$, and the 4--10\,keV count rate was 
$0.55\pm 0.02$\,s$^{-1}$\,cm$^{-2}$ (obtained from the {\em MAXI} 
website\footnote{http://maxi.riken.jp/top/}), demonstrating that the source 
was in the soft state based on the {\em MAXI} count rate and hardness criteria 
determined by \cite{grinberg13}.

\subsection{NuSTAR}

We reduced the data from the two {\em NuSTAR} instruments, Focal Plane Modules 
(FPMs) A and B, and the exposure times and other observation details are listed
in Table~\ref{tab:obs}.  The {\em NuSTAR} FPMs are Cadmium-Zinc-Telluride (CZT) 
pixel detectors with an energy resolution (full width at half-maximum) of 0.4\,keV 
at 10\,keV and 0.9\,keV at 68\,keV \citep{harrison13}.  Each FPM is at the focus of 
a hard X-ray telescope with a focal length of 10.14\,m and an angular resolution 
(half-power diameter) of $58^{\prime\prime}$ \citep{harrison13}.  We processed the 
{\em NuSTAR} data (ObsIDs 30001011002 and 30001011003) with version 1.1.1 of the 
NuSTARDAS pipeline software, the 2013 May 9 version of the {\em NuSTAR} Calibration 
Database (CALDB), and High Energy Astrophysics Software (HEASOFT) v6.13.  We 
produced cleaned event lists with the routine {\ttfamily nupipeline} and light curves 
and spectra with {\ttfamily nuproducts}.  The source extraction region is centered 
on Cyg~X-1 and has a radius of $200^{\prime\prime}$.  The background region is a 
$90^{\prime\prime}$ circle that is taken from the part of the {\em NuSTAR} field-of-view 
that is farthest from the source.  For ObsID 30001011002, the centers of the two 
regions are $10.\!^{\prime}5$ apart, and, for ObsID 30001011003, they are separated by 
$9.\!^{\prime}1$.  While the background rate is known to vary across the field-of-view
at low energies \citep{harrison13}, the source rate is 25--1000 times the background 
below 30\,keV, so systematic errors in the background cannot affect our results 
over this energy band. At higher energies (combining the energy bins above 30\,keV), 
the source is 21 times the background rate, so small detector-to-detector variations 
in the background are not important.

\subsection{Suzaku}

{\em Suzaku} covers the $\sim$0.3--600\,keV band via three detectors: the
X-ray Imaging Spectrometers \citep[XISs;][]{koyama07}, which are CCDs, the 
Hard X-ray Detector \citep[HXD;][]{takahashi:07a} PIN diode detector, and 
the HXD gadolinium silicate crystal detector (\gso).  These instruments
cover the $\sim$0.3--10\,keV, the $\sim$10--70\,keV, and the $\sim$60--600\,keV 
bands, respectively.  In this paper, for the XIS, we only consider the XIS0 
and XIS1 detectors.  During our observation, XIS3 was operated in a continuous 
readout mode (\texttt{PSUM} mode), complicating its analysis, while XIS2 has 
not been operational since 2006.

To create spectra from the {\em Suzaku} data (ObsID 407072010), we used tools 
from the HEASOFT v6.13 package and the calibration files current as of 
2013 February. We followed the standard procedure for analyzing the \xis spectra, 
which included correcting for Charge Transfer Inefficiency (CTI) and reprocessing 
the data with the \texttt{xispi} and \texttt{xselect} tools, respectively.  
Thermal bending of the spacecraft leads to attitude uncertainties, which in turn 
leads to distortions of the PSF image as observed by XIS. Although the standard 
HEASOFT tools apply corrections to the spacecraft attitude in order to improve 
the PSF image \citep{uchiyama:08a}, we further correct this image using the 
\texttt{aeattcor2} tool as described by \cite{nowak11}.

The \xis spectra were obtained in a mode where only 1/4 of the CCD was exposed 
with each CCD readout frame being 2\,s.  The spectra, however, were only exposed 
for 0.135\,s per readout frame in order to reduce telemetry and minimize pile-up.  
Despite these precautions, given the brightness of \cyg in its soft state, the 
spectra are heavily piled-up.  To estimate the degree of pile-up we employed the 
\texttt{pile\_estimate.sl} \texttt{S-Lang} script \citep[see][]{nowak11}.  Using 
this script, we identified the most heavily piled regions on the CCD and excluded 
two rectangular regions in the center each measuring approximately 
130$\times$45 pixels.  We estimate that the remaining regions on the \xis CCDs 
have an effective pile-up fraction of $\lesssim 5\%$.  We then used \texttt{xisrmfgen} 
and \texttt{xissimarfgen} to create response matrices for the extracted spectra.  
To account for systematics, we added a 2\% uncertainty on the \xis spectra in 
quadrature with the statistical uncertainties.

Standard procedures, following the \emph{Suzaku ABC
  Guide}\footnote{See
  http://heasarc.gsfc.nasa.gov/docs/suzaku/analysis/abc/}, were used
to create \hxd spectra.  \pin spectra were extracted from the
\texttt{hxd/event\_cl} directories with response and background files
downloaded from the \texttt{pinxb\_ver2.0\_tuned} directory at the
High Energy Astrophysics Science Archive Research Center
(HEASARC)\footnote{See http://heasarc.gsfc.nasa.gov/}.  \gso spectra
were created from ``unfiltered'' event files using the
\texttt{hxdtime}, \texttt{hxdpi}, and \texttt{hxdgrade} tools and the
filtering criteria from the standard \texttt{gso\_mkf.sel} script.
The background was obtained from the \texttt{gsonxb\_ver2.0} directory
at HEASARC. Event and background file Good Time Intervals were merged
to obtain the extraction times for the \gso spectra.  Standard CALDB
response files were applied to the spectra with their exposure times
adjusted to agree with the spectra. The grouping of the \gso spectra
followed the fixed grouping of the background file and thus were not
rebinned further.

\section{Results}

\begin{figure}
\includegraphics[scale=0.48]{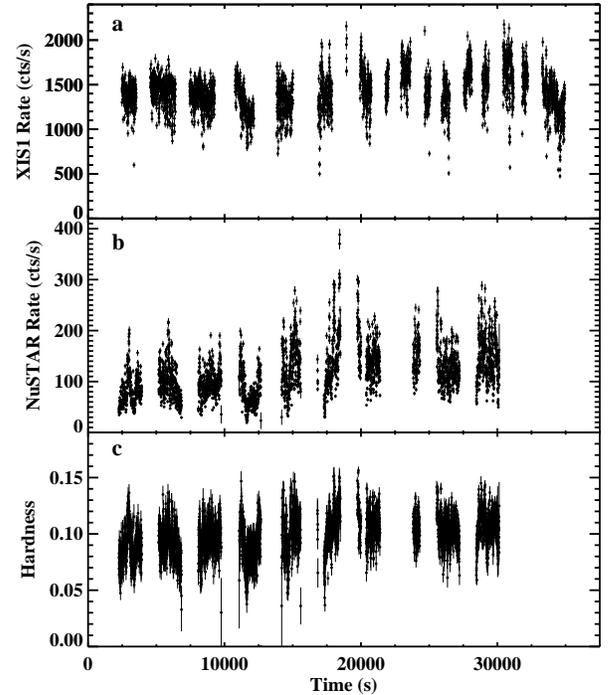}
\caption{\small {\em Suzaku}/XIS light curve {\em (a)}, {\em NuSTAR}/FPMA
light curve {\em (b)}, and {\em NuSTAR} hardness ratio {\em (c)} 
for Cyg~X-1.  For XIS, the bandpass is 0.5--9\,keV, and the rate is for 
XIS1 (after removing the piled-up core of the point spread function).  
For the {\em NuSTAR} light curve, the bandpass is 3--79\,keV, the 
rate is for FPMA, and it is corrected for deadtime.  The {\em NuSTAR} 
hardness ratio is the 10--79\,keV rate divided by the 3--10\,keV rate, 
and both modules are used.  The time resolution for all plots is 10\,s.  
The zero time is arbitrary but corresponds to MJD~56,231.30000.  For 
both satellites, most of the gaps are due to Earth occultation, but 
the longer gap near time 22,500\,s for {\em NuSTAR} is due to a missed 
ground station pass.\label{fig:lc}}
\end{figure}

Figure~\ref{fig:lc} shows the 3--79\,keV {\em NuSTAR} and 0.5--9\,keV XIS light 
curves.  There is good overlap in the coverage between the two satellites; however, 
their Earth occultations are not exactly in phase, and the {\em Suzaku} 
coverage extends somewhat beyond {\em NuSTAR}'s.  Flaring, which is typical of 
Cyg~X-1 in this state, is more evident in {\em NuSTAR}'s hard X-ray band than in 
the softer X-ray regime covered by XIS.  Perhaps the most notable feature in the 
XIS light curves are brief drops in the count rate.  It is possible that these 
are absorption dips due to material in the massive donor star's stellar wind.  
This is plausible because the observations occurred at a binary orbital phase of 
0.85-0.97 (where 1.0 corresponds to superior conjunction when the donor star is 
between the observer and the black hole) based on the ephemeris of \cite{brocksopp99}.  
Absorption dips are typically seen in this range of orbital phase \citep{bc00,poutanen08}.

In order to determine the level of spectral variation during the observations, 
we extracted the 3--10\,keV and 10--79\,keV {\em NuSTAR} count rates, and 
produced a plot of hardness, which is the 10--79\,keV count rate divided by
the 3--10\,keV count rate, vs.~time (Figure~\ref{fig:lc}c).  Even during the 
flares, we see little variability in the hardness.  Given the relatively low 
level of spectral variability, we combined all of the data into a single spectrum.

Due to the high count rate for Cyg~X-1, the XIS spectra show features that we 
suspect are related to photon pile-up.  An upturn in the spectra above $\sim$9\,keV 
is observed and is readily explained by pile-up.  The spectrum below 1.2\,keV shows 
features that appear to be absorption lines; however, we cannot rule out the 
possibility of some distortion due to instrumental effects, and we defer a detailed 
study to a later paper.  In addition, there are known calibration uncertainties in 
the 1.7--1.9\,keV band related to the Si K-edge.  After these considerations, for 
XIS0 and XIS1, we used the 1.2--1.7\,keV and 1.9--9\,keV bands for spectral analysis 
and binned the data based on the instrumental energy resolution 
(see Nowak et al.~2011\nocite{nowak11}).  For PIN and GSO, we used the 15--68\,keV 
and 50--296\,keV energy ranges, respectively.  For {\em NuSTAR}, we used 3--79\,keV, 
and binned the spectra for FPMA and FPMB separately, requiring that each bin have 
a signal-to-noise ratio of at least 30 (after background subtraction).

\begin{figure}
\includegraphics[scale=0.36]{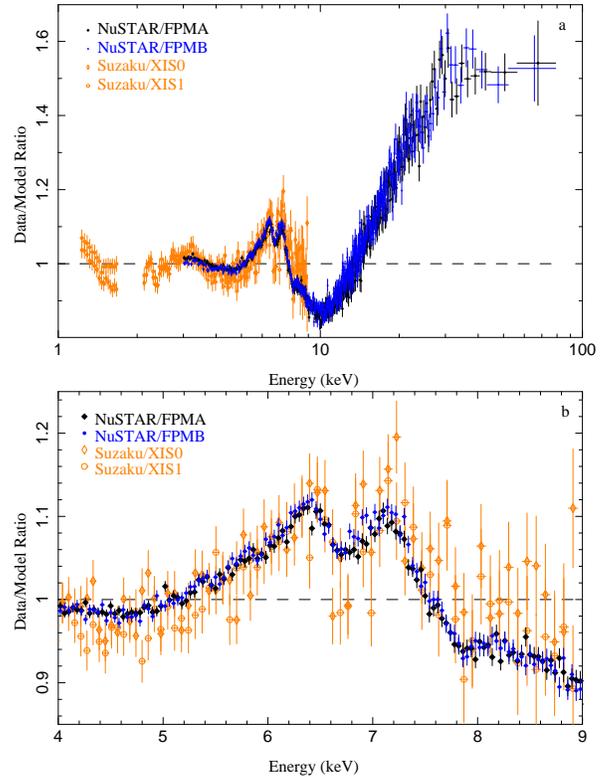}
\caption{\small The data-to-model ratio for a fit to the Cyg~X-1 spectrum 
with an absorbed disk-blackbody plus power-law model (model 1).  The panel 
{\em (a)} residuals indicate a strong reflection component.  Panel {\em (b)} 
focuses on the iron K$\alpha$ line region, showing that the line complex 
includes at least a broad emission component and an absorption line at 6.7\,keV.
\label{fig:ratio}}
\end{figure}

We used the XSPEC software package \citep{arnaud96} to fit the combined 
{\em NuSTAR} plus {\em Suzaku} spectrum with a model consisting of a 
multi-temperature ``disk-blackbody'' thermal component \citep{mitsuda84} 
plus a power-law (model 1).  These continuum components were subject to absorption 
with the {\ttfamily tbabs} model, and we used \cite{wam00} abundances and 
\cite{vern96} cross-sections for this interstellar absorption.  We included a 
multiplicative constant as a free parameter for each instrument to account for 
differences in overall normalization.  Figure~\ref{fig:ratio} shows the XIS and 
{\em NuSTAR} residuals for this fit in terms of the data-to-model ratio, revealing 
a strong reflection component with a broad iron K$\alpha$ emission line and a 
reflection hump above $\sim$15\,keV.  Figure~\ref{fig:ratio}b illustrates the 
complexity of the iron line, which has an absorption line at 6.7\,keV in addition 
to the broad line in emission.  

The fit can be significantly improved with the addition of a Gaussian emission
line and a cutoff at high energies (model 2), using {\ttfamily highecut}, which 
provides an exponential cutoff with a folding energy of $E_{\rm fold}$ for energies 
greater than a cutoff energy, $E_{\rm cut}$.  If the Gaussian parameters are allowed 
to take any values, the line centroid is near 5.3\,keV, which is well below the
6.4--7.1\,keV iron regime, and the line is extremely broad ($\sigma = 1.57$\,keV).
In addition to the Gaussian parameter values being unphysical, this model does 
not give a formally acceptable fit with a reduced-$\chi^{2}$ ($\chi^{2}_{\nu}$) of 
2.00 for 1149 degrees of freedom (dof).  The continuum parameters (e.g., a best 
fit inner disk temperature of $kT_{\rm in} = 0.62$\,keV and a power-law photon 
index of $\Gamma = 2.5$) are consistent with the source being in the soft state.  
This model gives absorbed and unabsorbed 0.5--100\,keV fluxes of 
$4.33\times 10^{-8}$\,erg~cm$^{-2}$~s$^{-1}$ and 
$6.09\times 10^{-8}$\,erg~cm$^{-2}$~s$^{-1}$, respectively.  For a source distance
of 1.86\,kpc, this implies a luminosity of $2.5\times 10^{37}$\,erg~s$^{-1}$, which, 
for a BH mass of 14.8\Msun, gives an Eddington-scaled luminosity of 1.3\%.

\begin{figure}
\hspace{-0.5cm}
\includegraphics[scale=0.48]{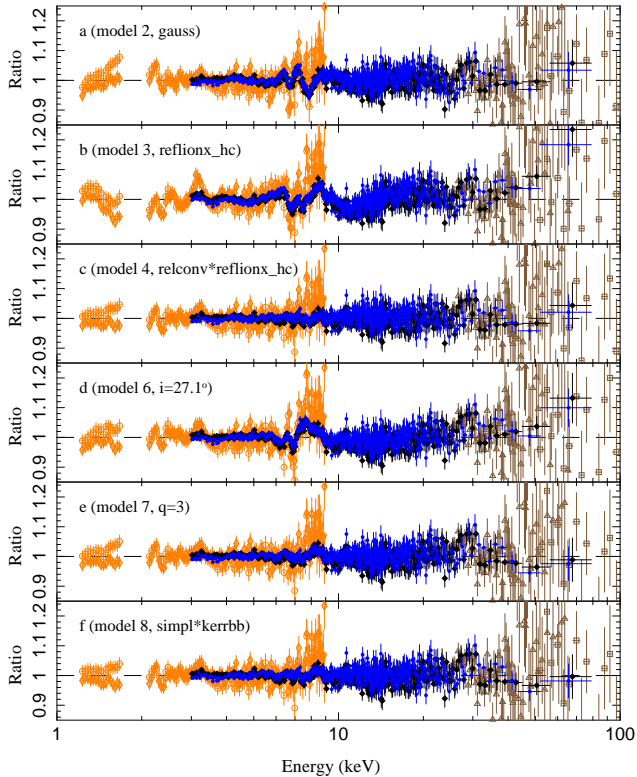}
\vspace{-0.4cm}
\caption{\small Data-to-model ratios for six of the models described in 
\S\,3.  The symbols and colors for XIS0, XIS1, FPMA, and FPMB are
the same as for Figure~\ref{fig:ratio}.  In addition, at higher energies, 
the PIN and GSO ratios, which are only shown up to 100\,keV, are marked 
with brown triangles and squares, respectively.\label{fig:ratio_many}}
\end{figure}

As shown in Figure~\ref{fig:ratio_many}a, the largest residuals for model 2 are 
in the 6--8.5\,keV part of the spectrum.  In addition to the fact that we are still 
not modeling the 6.7\,keV absorption line, which is due to the photoionized wind
of the massive companion star, a Gaussian is too simple to fit the broad emission 
feature and the absorption edge that are present in the reflection component.  
Thus, we removed the Gaussian and added a simple ionized absorber and a reflection 
component (model 3).  

For absorption due to the wind, we constructed a grid of table models using XSTAR 
version 2.2.1bg \citep{kb01}.  Solar abundances were assumed for all elements,
the number density was fixed at $n = 10^{12}~ {\rm cm}^{-3}$, and the turbulent 
velocity of the gas was fixed at $v_{\rm turb} = 300~ {\rm km}~ {\rm s}^{-1}$
\citep[e.g.,][]{miller05,hanke09}.  We used an input spectrum consistent with 
model 1 described above in order to construct a grid spanning 
$2 \leq {\rm log}(\xi) \leq 5$, where $\xi$ is the ionization parameter in 
units of erg\,cm\,s$^{-1}$, and 
$1.0\times 10^{21}~ {\rm cm}^{-2} \leq N_{\rm H} \leq 5.0\times 10^{22}~ {\rm cm}^{-2}$, 
where $N_{\rm H}$ is the column density of the absorber.  In total, 400 grid points 
were calculated and summed into a multiplicative table model that was included in 
XSPEC analysis, with $N_{\rm H}$, $\xi$ and $v/c$ as variable parameters.  Although 
$v/c$ was originally left as a free parameter, we found a 90\% confidence upper limit 
of $<$0.0004, and we fixed it to zero in the fits described below.  This parameter is 
driven by the strong absorption line at 6.7\,keV, which is due to Fe XXV.

For the reflection, we used the {\ttfamily reflionx} model \citep{rf05}.  This 
model includes the hard X-ray bump, the absorption edges, and the emission lines, 
so that the full reflection component is physically self-consistent.  In addition, 
the emission lines are Compton-broadened (see \S\,1).  The version that is available
on-line\footnote{See http://heasarc.gsfc.nasa.gov/xanadu/xspec/models/reflion.html.} 
has the folding energy for its exponential cutoff fixed at 300\,keV, but, for
our fits, a new model, {\ttfamily reflionx\_hc}, was produced with $E_{\rm fold}$ as a 
free parameter.  For the direct component, the {\ttfamily highecut} parameters were 
set to be consistent with {\ttfamily reflionx\_hc}:  $E_{\rm cut}$ was set to zero and 
$E_{\rm fold}$ was forced to have the same value as the free parameter in the 
reflection model.  One other difference between {\ttfamily reflionx} and 
{\ttfamily reflionx\_hc} is that the ionization parameter was extended to higher 
levels based on early fits to the Cyg~X-1 spectrum.  While this is a more 
realistic physical model than using the Gaussian to fit the iron line, model 3 
provides a worse fit ($\chi^{2}_{\nu}$=2.72 for 1148 dof) than model 2, and 
large residuals are still present in the 5--9\,keV regime.

\begin{figure}
\hspace{-0.3cm}
\includegraphics[scale=0.43]{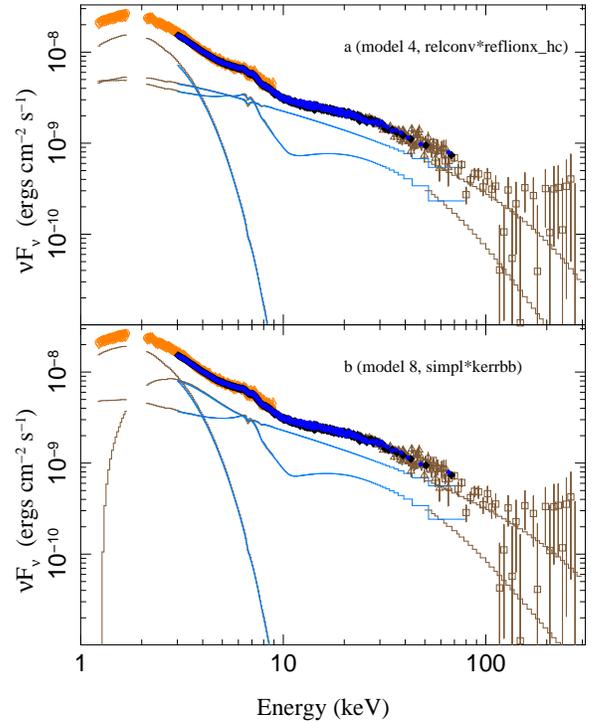}
\vspace{-0.5cm}
\caption{\small {\em (a)} Unfolded {\em NuSTAR} and {\em Suzaku} spectrum showing 
the fit obtained with model 4, which includes a disk-blackbody component, a 
cutoff power-law, a {\ttfamily reflionx\_hc} reflection model with relativistic 
blurring, and a simple ionized absorber. {\em (b)} The spectrum for model 8, which 
models the thermal component with {\ttfamily kerrbb} and is self-consistent in that 
the thermal component is the seed photon distribution for the Comptonized component
(using {\ttfamily simpl}).  The symbols and colors for the different instruments are 
the same as for Figure~\ref{fig:ratio_many}.\label{fig:efe}}
\end{figure}

A major improvement in the fit (to $\chi^{2}_{\nu}$=1.21 for 1143 dof) comes from 
convolving the reflection component with a relativistic blurring model (model 4).  
For blurring, we used the {\ttfamily relconv} model \citep{dauser10}, which is 
based on the physics described in \cite{fabian89} and \cite{laor91}, but 
{\ttfamily relconv} allows for a range of spin values.  For these fits, we assume 
that the accretion disk extends to the ISCO, and the blurring, which is most 
apparent in its effect on the iron line shape, depends on the BH spin ($a_{*}$), 
the disk inclination ($i$), and the radial dependence of the emissivity of reflected 
flux.  The emissivity is assumed to have a power-law ($L\propto r^{-q}$, where $L$ 
is the luminosity illuminating the reflecting material, $r$ is the radial distance 
from the BH, and $q$ is the emissivity index) or broken power-law shape.  The fit 
parameter values for the broken power-law emissivity (model 4) and for the power-law 
emissivity (model 5) are given in Table~\ref{tab:parameters1}, and 
Figure~\ref{fig:efe}a shows the components of the former model.  

\begin{figure*}
\includegraphics[scale=0.55]{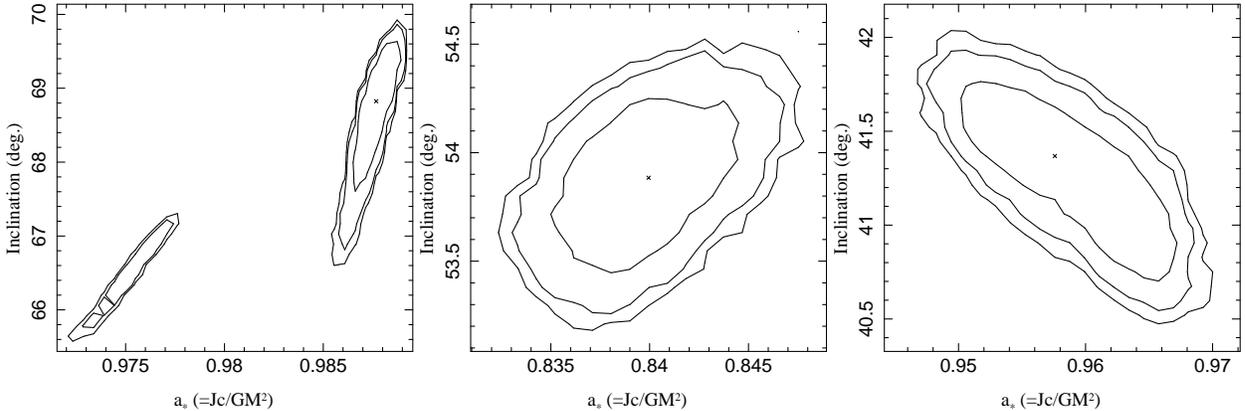}
\vspace{-0.5cm}
\caption{\small Error contours for BH spin and inner disk inclination for
models 4 {\em (left)}, 8 {\em (middle)}, and 10 {\em (right)}.  The 1, 2, 
and 3-$\sigma$ contours are shown.
\label{fig:contours}}
\end{figure*}

The fit parameters indicate reflection off highly ionized material 
($\xi$$>$13,900\,erg\,cm\,s$^{-1}$ for model 4 and $\xi$$>$19,500\,erg\,cm\,s$^{-1}$ 
for model 5) and a steep emissivity index ($q$$>$9.5).  In addition, we find a 
high BH spin of $a_{*} = 0.9882\pm 0.0009$ for model 4 and 
$a_{*} = 0.91^{+0.01}_{-0.02}$ for model 5.  These are 90\% confidence statistical 
errors, and it is important to note that they do not include any systematic 
component.  The inclinations obtained are $i = 69.2^{+0.5}_{-0.9}$~degrees and 
$i = 59.3^{+0.5}_{-1.3}$~degrees for models 4 and 5, respectively, both of which 
are significantly different from the value of 27.1 degrees measured for the 
binary \citep{orosz11}.  If we fix the inclination to the binary value and 
refit the spectrum, we obtain a very poor fit with $\chi^{2}_{\nu}$=2.45 for 
1144 dof even for the case of broken power-law emissivity, and the residuals 
are shown in Figure~\ref{fig:ratio_many}d (model 6).  Furthermore, we made
error contours\footnote{To explore correlations among parameters, we performed 
Markov Chain Monte Carlo (MCMC) simulations with a code modeled after the ``emcee
hammer'' code described by \cite{fm13}, which  implements the algorithm of 
\cite{gw10}.  In this algorithm, an ensemble of ``walkers,'' which are vectors 
of the fit parameters, are evolved via random steps determined by the difference 
between two walkers.  We evolved 20 walkers per free parameter for a total of 
4000--10,000 steps, and ignored the first half of the steps.  Thus, probability 
distributions were calculated from (0.4--1.5)$\times 10^{6}$ values.  Error 
contours are the 2D projection of the MCMC N-dimensional probability distribution.}
for the spin and inclination parameters for model 4 (see Figure~\ref{fig:contours}).  
Although there is some correlation between these parameters, and a nearby local 
minimum exists, the 3-$\sigma$ contours do not extend below $i$$\sim$$65.8$ degrees.

For all the models presented thus far, if $q_{\rm in}$ is left as a free parameter, 
we obtain values close to 10, which is the maximum of the allowed range.  We also 
explored the implications of lower emissivity index by fixing it to $q = 3$.  
This gives $\chi^{2}_{\nu}$=1.40 for 1146 dof, which is significantly worse than 
the high-$q$ (and high-$i$) fit, but the inclination is $42.4^{+0.4}_{-0.5}$~degrees, 
which is much closer to the binary value.  The residuals for this model are shown 
in Figure~\ref{fig:ratio_many}e (model 7), and the parameters are given in 
Table~\ref{tab:parameters1}.  Although the BH spin is somewhat lower for model 7, 
the relatively poor fit suggests that this value is not reliable.  For model 4, we 
left $q_{\rm out}$ as a free parameter, and a value of --$1.2^{+1.1}_{-4.6}$ is obtained, 
indicating that, beyond 10\,$R_{\rm g}$, the flux incident on the disk is actually 
increasing with radius.  Although such a rising profile could occur over some range 
of radii, we note that it is non-physical for the emissivity to continue to 
increase with radius indefinitely.

While the parameter and BH spin constraints above rely only on modeling 
the reflection component, a previous Cyg~X-1 spin measurement obtained by 
fitting the soft state spectrum relied primarily on modeling the thermal 
component \citep{gou11,gou13}.  Rather than using the disk-blackbody model, 
they used the model {\ttfamily kerrbb}, which is a multi-temperature thermal 
accretion disk model that accounts for changes in the inner disk (e.g., the 
inner radius) due to the BH spin.  Also, instead of adding a power-law, they 
used the convolution model {\ttfamily simpl} \citep{steiner09}, which is 
different from the disk-blackbody plus power-law model described above 
because it uses the {\ttfamily kerrbb} component as the seed photon input to 
the Comptonization region.  With this model, we obtain $\chi^{2}_{\nu}$ = 1.32 
for 1146 dof.  The residuals are shown in Figure~\ref{fig:ratio_many}f 
(model 8), and the model components are shown in Figure~\ref{fig:efe}b.  
The parameter values from the fit are given in Table~\ref{tab:parameters2}.  
The constraint on the spin parameter, $a_{*} = 0.838\pm 0.006$, comes 
from both the the thermal component and the reflection component, and the 
inclination ($i = 53.9\pm 0.4$~degrees) is still significantly higher than 
the binary value.  Model 8 uses a single power-law for the emissivity with an 
index of $q = 7.8\pm 0.5$.  Figure~\ref{fig:contours} shows the error 
contours for spin and inclination for model 8.

Although we do not focus on calibration details in this paper, there is 
excellent agreement between FPMA and FPMB with the relative normalization 
being consistent to within 0.1\% for all the spectral models described 
above, which is actually better than expected.  Relative to {\em NuSTAR}/FPMA, 
we find normalization constants of $1.081\pm 0.005$ for XIS0, $1.038\pm 0.004$ 
for XIS1, $1.205\pm 0.007$ for PIN, and $1.17\pm 0.06$ for GSO.  
These numbers are for model 8, but Tables~\ref{tab:parameters1} and
\ref{tab:parameters2} show very similar relative normalizations for all
models.  Thus, there is very good agreement between {\em NuSTAR} and
XIS, and the fact that PIN and GSO are somewhat higher is 
expected\footnote{See http://www.astro.isas.jaxa.jp/suzaku/doc/suzakumemo/suzakumemo-2008-06.pdf}.

\section{Discussion}

The combination of {\em NuSTAR} and {\em Suzaku} provide a measurement of 
the Cyg~X-1 reflection spectrum with unprecedented quality.  While {\em NuSTAR} 
measures the entire reflection component (iron line, absorption edges, and 
hard X-ray bump), the XIS provides an extension to lower energies that is 
essential for constraining the thermal component.  {\em NuSTAR} and XIS agree 
to a remarkable extent on the shape of the iron line (see Figure~\ref{fig:ratio}b), 
and {\em NuSTAR} provides a huge improvement in the statistical quality of the 
data, while alleviating some systematic concerns such as pile-up.  

We have presented fits to the spectrum with several different models, 
and, while some parameters show significant differences, others agree
about the properties of the system.  It is clear that the source was
in the soft state with a prominent thermal component and a power-law
with a photon index between $\Gamma = 2.59$ and 2.67, which meets 
the $\Gamma > 2.5$ criterion for Cyg~X-1 to be in the soft state
\citep{grinberg13}.  There is clear evidence for absorption due to 
highly ionized material, which is consistent with the findings of 
\cite{yamada13}.  Also, the fits agree that the ionization state of 
the disk material that leads to the reflection component is high and 
that iron is overabundant by a factor of 1.9--2.9 relative to solar.  

The BH spin and inclination measurements vary from model-to-model by more 
than the 90\% confidence statistical errors, indicating that there is 
significant systematic uncertainty.  For the inclination, it is also 
necessary to compare our values of $i = 42$--69 degrees to the 
value of $i = 27.1\pm 0.8$~degrees that is obtained by modeling optical 
photometric and spectroscopic measurements \citep{orosz11}, but the optical 
measurement is of the inclination of the binary while the reflection component 
measures the inclination of the inner part of the disk.  We have shown (see 
Figure~\ref{fig:ratio_many}d) that the reflection model simply cannot reproduce 
an inclination as low as 27.1 degrees.

However, we must also keep the limitations of the spectral model in mind.  
While the {\ttfamily relconv} calculation is for a specific inclination angle, 
{\ttfamily reflionx\_hc} calculates the spectrum of the reflection component by 
averaging over angles.  Another consideration is that the ionization parameter 
($\xi$) is at the top of the available range for {\ttfamily reflionx\_hc}.  The 
model already includes Compton broadening of the lines, which increases with 
increasing $\xi$, but it is possible that some extra Compton broadening is 
necessary to account for a higher ionization.  Also, surface turbulence may 
cause some symmetric line broadening that is not taken into account by 
{\ttfamily reflionx\_hc}.  To test this, we added a Gaussian convolution model 
({\ttfamily gsmooth}), which acts on the reflection component along with 
{\ttfamily relconv}.  The results are shown in Table~\ref{tab:parameters2}, 
where this is listed as model 9, and the inclination decreases significantly 
from 53.9 to 40.4 degrees.  The value obtained for the {\ttfamily gsmooth} 
$\sigma$ parameter is $0.28^{+0.02}_{-0.04}$\,keV.  Using 
$kT = (1/2)m_{e}c^{2}(\sigma/E)^{2}$, where $m_{e}$ is the electron rest mass, 
this value of $\sigma$ corresponds to a temperature of $kT = 0.4$--0.6\,keV at 
$E = 6$--7\,keV, which is in-line with the inner disk temperatures we obtain 
from the disk-blackbody fits.

While adding symmetric smoothing of the iron line and reflection component
(i.e., extra Compton broadening) causes a drop in $i$, we emphasize that 
asymmetric relativistic broadening is required by the data.  For our original 
Gaussian fit to the iron line, we obtained a best fit centroid value of 5.3\,keV, 
which shows that the line has the low-energy tail expected for a gravitational 
redshift.  Also, we obtained a very poor fit with {\ttfamily reflionx\_hc} 
(Figure~\ref{fig:ratio_many}b), where the Compton broadening was already included.  
Adding the relativistic broadening provided a very large improvement to the fit 
(Figure~\ref{fig:ratio_many}c).

Although conclusions about the BH spin depend on the different possibilities
for the inclination, the models which provide good fits to the data
(models 4, 5, 8, and 9) all have $a_{*}$$>$0.83, indicating at least relatively
high spin.  The best fit (model 4) also has the highest spin
$a_{*} = 0.9882\pm 0.0009$, but this either requires a very large warp
in the accretion disk or that the binary inclination is somewhat higher
than the best fit value found in \cite{orosz11}.  We note that Table 1
in \cite{orosz11} reports that some of their models give significantly 
higher inclinations, but the $\chi^{2}$ values for the higher inclination 
models are worse.  

Another potentially interesting result that comes from this spectrum is 
the constraint on the emissivity profile.  A comparison of models 4--6 
indicate that a broken power-law emissivity is preferred as is a very steep 
profile in the inner part of the disk ($q_{\rm in}$$\sim$10).  For Active
Galactic Nuclei (AGN), relatively steep profiles ($q = 4.3$--5.0) were 
reported for MCG--6-30-15 \citep{wilms01}, and steep profiles are discussed 
in \cite{wf12}.  \cite{walton13} studied a large sample of AGN, and found 
that steep profiles are common.  This has been taken as evidence that the 
irradiating source comes from very close to the BH, and \cite{fabian12a} 
conclude that it must lie within 1\,$R_{\rm g}$ of the BH event horizon.  
While this may also be the case for Cyg~X-1, our fits with very steep 
profiles (models 4, 5, and 8) also have inclinations between 53.9 and 
69.2 degrees.  Our fit with {\ttfamily gsmooth} (model 9) included in the 
model gave a much flatter index of $q = 2.48^{+0.09}_{-0.05}$, leaving open 
the possibility that the profile is relatively flat, in which case the 
source is at a height of 5--10\,$R_{\rm g}$ or more.

While we cannot conclude anything definitive about the slope of the emissivity
profile, if it is very steep, this might point to a ``lamppost'' geometry 
\citep{dauser10}, where the emission actually comes from the base of a 
collimated jet.  This geometry may not be relevant for the soft state 
because there is no evidence for a jet.  Despite this, if we start with
model 8 but replace {\ttfamily relconv} with {\ttfamily relconv\_lp}
(model 10), we find $i = 41.5\pm 0.5$ degrees and $a_{*} = 0.953\pm 0.006$
(see Figure~\ref{fig:contours} for the error contours), with only small 
changes in the other parameters.  However, the quality of the fit is 
somewhat worse ($\chi^{2}_{\nu} = 1.44$) for model 10 compared to the models 
reported in Tables~\ref{tab:parameters1} and \ref{tab:parameters2}.

After exploring different continuum models, emissivity geometries, and 
conditions for the material in the accretion disk, we only find inner
disk inclinations that are $>$13 degrees higher than the binary value
measured by \cite{orosz11}, and, as mentioned above, one explanation is 
that there is a warp in the accretion disk. Analytical calculations as 
well as numerical simulations have shown that disk warps can occur 
\citep{bp75,sm94,fragile07}, and that they should occur if the BH spin 
is misaligned from the orbital plane (and outer disk).  As the alignment 
time for an accreting BH can be longer than the lifetime of a high-mass 
system \citep{maccarone02}, if the Cyg~X-1 BH formed with a misaligned spin, 
it would remain misaligned.  If jets are aligned with the BH spin, then
there is evidence for misalignment in systems like Cyg~X-3, V4641~Sgr, 
and GRO~J1655--40 \citep{maccarone02}.  It should be noted that 
\cite{fragile09} has shown that, under certain assumptions about the 
thickness of the accretion disk, BH spin measurements using the inner 
radius of a warped disk can be incorrect.  While a disk warp may not be 
the only possibility for Cyg~X-1, further investigations to determine 
if the disk is really warped have important implications for the BH spin 
measurement.

\section{Summary and Conclusions}

We have presented a detailed study of the $\sim$1--300\,keV spectrum 
of Cyg~X-1 in the soft state.  The spectrum is complex and consists 
of multi-temperature blackbody, power-law, and reflection components 
along with absorption from highly ionized material in the system.  
Although the observation was of moderate duration ($\sim$29\,ks), 
{\em NuSTAR} provides a very high-quality and high-statistics measurement 
of the reflection spectrum, including an iron complex with broad emission 
and narrow absorption lines.  We find that the reflecting material 
has a high ionization state, is overabundant in iron relative to 
solar, and requires broadening of the iron line that is well-described 
by a relativistic blurring model.  

While all models that provide a good fit to the spectrum indicate a
rapidly rotating BH with $a_{*}$$>$0.83, and our best-fitting model 
has $a_{*} = 0.9882\pm 0.0009$ (90\% confidence statistical
errors only), we were not able to obtain a good fit with the inclination 
fixed to the \cite{orosz11} binary value.  This may indicate a 
misalignment between the orbital plane and the inner accretion disk
(by $>$13 degrees), missing physics in the spectral models, or it may 
possibly motivate work to confirm the measurement of the binary inclination.  
Regardless of which of these possibilities is correct, it is clear that 
the combination of {\em NuSTAR}'s high throughput and energy resolution 
provides a major advance in reflection studies, allowing for strict tests 
of the models, which we expect to lead to improved constraints on the 
physical processes at work in Cyg~X-1 and other accreting BH systems.

\acknowledgments

This work was supported under NASA Contract No. NNG08FD60C, and made use of data 
from the {\it NuSTAR} mission, a project led by  the California Institute of 
Technology, managed by the Jet Propulsion  Laboratory, and funded by the National 
Aeronautics and Space Administration. We thank the {\it NuSTAR} Operations, 
Software and  Calibration teams for support with the execution and analysis of 
these observations.  This research has made use of the {\it NuSTAR}  Data 
Analysis Software (NuSTARDAS) jointly developed by the ASI  Science Data 
Center (ASDC, Italy) and the California Institute of  Technology (USA).
JAT acknowledges partial support from NASA Astrophysics Data Analysis Program 
grant NNX13AE98G.  LN wishes to acknowledge the Italian Space Agency (ASI) for 
financial support by ASI/INAF grant I/037/12/0-011/13.  JAT thanks L.~Brenneman, 
G.~Matt, and D.~Ballantyne for useful discussions about reflection modeling.  
This work made use of IDL software written by N.~Barri{\`e}re for rebinning the 
{\em NuSTAR} spectra.  This research has made use of the {\em MAXI} data provided 
by RIKEN, JAXA, and the {\em MAXI} team.


\clearpage


\begin{table}
\caption{Observing Log and Exposure Times\label{tab:obs}}
\begin{minipage}{\linewidth}
\begin{center}
\begin{tabular}{ccccccc} \hline \hline
        &            &       & Start Time (UT) & End Time (UT) & On-Source & Exposure\\
Mission & Instrument & ObsID & (in 2012)       & (in 2012)     & Time (ks) & (s)\\
\hline\hline
{\em NuSTAR} & FPMA     & 30001011002 & Oct 31, 8.18 h  & Oct 31, 17.77 h & 18.4 & 10,442\\
{\em NuSTAR} & FPMB     &   ''        &       ''        &           ''    &  ''  & 10,811\\
{\em NuSTAR} & FPMA     & 30001011003 & Oct 31, 17.77 h & Nov 1, 0.27 h   & 10.3 &  5,096\\
{\em NuSTAR} & FPMB     &      ''     &       ''         &          ''    &  ''  &  5,257\\
{\em Suzaku} & XIS0     & 407072010   & Oct 31, 8.20 h  & Nov 1, 2.62 h   & 30.1 &  1,939\\
{\em Suzaku} & XIS1     &      ''     &       ''         &          ''    &  ''  &  1,991\\
{\em Suzaku} & HXD/PIN  &      ''     &       ''         &          ''    &  ''  & 30,074\\
{\em Suzaku} & HXD/GSO  &      ''     &       ''         &          ''    &  ''  & 27,880\\ \hline
\end{tabular}
\end{center}
\end{minipage}
\end{table}

\begin{table}
\caption{Fit parameters for models with disk-blackbody\label{tab:parameters1}}
\begin{minipage}{\linewidth}
\begin{center}
\begin{tabular}{ccccc} \hline \hline
Parameter & Unit/Description & Model 4 Value\footnote{With 90\% confidence errors.  A value of zero for the positive error indicates that the parameter's error range reached the upper limit of values provided for the model.} & Model 5 Value$^{a}$ & Model 7 Value$^{a}$\\ \hline
\multicolumn{5}{l}{Interstellar Absorption}\\
$N_{\rm H}$ & $10^{21}$\,cm$^{-2}$ & $6.0\pm 0.3$ & $6.2\pm 0.2$ & $6.2\pm 0.2$\\ \hline
\multicolumn{5}{l}{Disk-blackbody}\\
$kT_{\rm in}$ & keV & $0.558^{+0.004}_{-0.002}$ & $0.558\pm 0.003$ & $0.557^{+0.004}_{-0.002}$\\
$N_{\rm DBB}$ & Normalization & $20,800^{+1200}_{-800}$ & $19,600^{+800}_{-600}$ & $19,600^{+1000}_{-900}$\\ \hline
\multicolumn{5}{l}{Cutoff Power-law}\\
$\Gamma$ & Photon Index & $2.589^{+0.005}_{-0.022}$ & $2.66\pm 0.02$ & $2.672\pm 0.014$\\
$N_{\rm pl}$ & Normalization\footnote{In units of ph\,s$^{-1}$\,cm$^{-2}$\,keV$^{-1}$ evaluated at 1\,keV.} & $6.0\pm 0.4$ & $6.8^{+0.4}_{-0.2}$ & $7.4\pm 0.3$\\
$E_{\rm fold}$ & keV & $120^{+20}_{-10}$ & $190^{+20}_{-10}$ & $200^{+50}_{-20}$\\ \hline
\multicolumn{5}{l}{Simple Ionized Absorber}\\
$N_{\rm H}$ & $10^{22}$\,cm$^{-2}$ & $3.45^{+0.14}_{-0.23}$ & $3.31^{+0.29}_{-0.10}$ & $2.86\pm 0.17$\\ 
$\log{\xi}$ & erg\,cm\,s$^{-1}$ & $5.0^{+0.0}_{-0.2}$ & $4.84^{+0.16}_{-0.02}$ & $5.00^{+0.00}_{-0.03}$\\ \hline
\multicolumn{5}{l}{Reflection Component ({\ttfamily reflionx\_hc})}\\
$\xi$ & erg\,cm\,s$^{-1}$ & $18,100^{+1900}_{-4200}$ & $20,000^{+0}_{-500}$ & $20,000^{+0}_{-800}$\\
Fe/solar & Abundance & $2.9\pm 0.4$ & $1.9\pm 0.2$ & $1.93^{+0.13}_{-0.23}$\\
$N_{\rm ref}$ & Normalization ($\times$$10^{-6}$) & $5.955^{+0.003}_{-0.327}$ & $6.6^{+0.5}_{-0.4}$ & $6.0^{+0.5}_{-0.3}$\\ \hline
\multicolumn{5}{l}{Relativistic Blurring ({\ttfamily relconv}\footnote{Two other parameters in this model are the inner and outer radii from where the reflected emission is coming: $R_{\rm in}$ is set to be at the ISCO; and $R_{\rm out} = 400 R_{\rm g}$.})}\\
$q_{\rm in}$ & Emissivity Index & $10.0^{+0.0}_{-0.4}$ & $10.0^{+0.0}_{-0.5}$ & 3.0\footnote{Fixed.}\\
$q_{\rm out}$ & Emissivity Index & --$1.2^{+1.1}_{-4.6}$ & $10.0$ & 3.0$^{c}$\\
$R_{\rm break}$ & Index Break Radius ($R_{\rm g}$) & $10^{+15}_{-3}$ & -- & --\\
$a_{*}$ & Black Hole Spin & $0.9882\pm 0.0009$ & $0.91^{+0.01}_{-0.02}$ & $0.75\pm 0.05$\\
$i$ & Inclination (degrees) & $69.2^{+0.5}_{-0.9}$ & $59.3^{+0.5}_{-1.3}$ & $42.4\pm 0.5$\\ \hline
\multicolumn{5}{l}{Cross-Normalization Constants (relative to FPMA)}\\
$C_{\rm XIS0}$ & -- & $1.081\pm 0.005$ & $1.082\pm 0.005$ & $1.081\pm 0.005$\\
$C_{\rm XIS1}$ & -- & $1.038\pm 0.005$ & $1.039\pm 0.005$ & $1.038\pm 0.005$\\
$C_{\rm FPMB}$ & -- & $1.001\pm 0.001$ & $1.001\pm 0.001$ & $1.001\pm 0.001$\\
$C_{\rm PIN}$ & -- & $1.202\pm 0.006$ & $1.204\pm 0.006$ & $1.207\pm 0.007$\\
$C_{\rm GSO}$ & -- & $1.23\pm 0.06$ & $1.17\pm 0.05$ & $1.16\pm 0.05$\\ \hline
$\chi^{2}/\nu$ & -- & 1388/1143 & 1501/1145 & 1610/1146\\ \hline
\end{tabular}
\end{center}
\end{minipage}
\end{table}

\begin{table}
\caption{Fit parameters for models with {\ttfamily kerrbb}\label{tab:parameters2}}
\begin{minipage}{\linewidth}
\begin{center}
\begin{tabular}{cccc} \hline \hline
Parameter & Unit/Description & Model 8 Value\footnote{With 90\% confidence errors.  A value of zero for the positive error indicates that the parameter's error range reached the upper limit of values provided for the model.} & Model 9 Value$^{a}$\\ \hline
\multicolumn{4}{l}{Interstellar Absorption}\\
$N_{\rm H}$ & $10^{21}$\,cm$^{-2}$ & $6.6\pm 0.2$ & $6.5\pm 0.1$\\ \hline
\multicolumn{4}{l}{Thermal Component ({\ttfamily kerrbb}\footnote{Fixed parameters and their values include $\eta = 0.0$, which corresponds to the zero torque inner boundary condition, $M_{\rm BH} = 14.8$\Msun, $d = 1.86$\,kpc, and a spectral hardening factor of 1.7.})}\\
$i$ & Inclination (degrees) & $53.9\pm 0.4$ & $40.4\pm 0.5$\\
$a_{*}$ & Black Hole Spin & $0.838\pm 0.006$ & $0.973\pm 0.004$\\
$\dot{M}$ & Accretion Rate ($10^{18}$\,g\,s$^{-1}$) & $0.203^{+0.004}_{-0.006}$ & $0.127^{+0.005}_{-0.007}$\\ \hline
\multicolumn{4}{l}{Comptonization ({\ttfamily simpl})}\\
$\Gamma$ & Photon Index & $2.66\pm 0.02$ & $2.65\pm 0.02$\\
$f_{\rm scat}$ & Scattering Fraction & $0.104\pm 0.004$ & $0.10499^{+0.00344}_{-0.00001}$\\
$E_{\rm fold}$ & keV & $190^{+40}_{-20}$ & $180^{+30}_{-10}$\\ \hline
\multicolumn{4}{l}{Simple Ionized Absorber}\\
$N_{\rm H}$ & $10^{22}$\,cm$^{-2}$ & $3.46^{+0.12}_{-0.20}$ & $3.27^{+0.08}_{-0.41}$\\ 
$\log{\xi}$ & erg\,cm\,s$^{-1}$ & $5.00^{+0.00}_{-0.13}$ & $5.0^{+0.0}_{-0.2}$\\ \hline
\multicolumn{4}{l}{Reflection Component ({\ttfamily reflionx\_hc})}\\
$\xi$ & erg\,cm\,s$^{-1}$ & $20,000^{+0}_{-1200}$ & $19,200^{+800}_{-3700}$\\
Fe/solar & Abundance & $1.99^{+0.11}_{-0.22}$ & $2.00^{+0.18}_{-0.14}$\\
$N_{\rm ref}$ & Normalization ($\times$$10^{-6}$) & $6.45^{+0.01}_{-0.23}$ & $5.9^{+1.3}_{-0.3}$\\ \hline
\multicolumn{4}{l}{Relativistic Blurring ({\ttfamily relconv}\footnote{Two other parameters in this model are the inner and outer radii from where the reflected emission is coming: $R_{\rm in}$ is set to be at the ISCO; and $R_{\rm out} = 400 R_{\rm g}$.  The inclination and spin parameters ($i$ and $a_{*}$) are free, but they are forced to take the same values as for {\ttfamily kerrbb}.})}\\
$q$ & Emissivity Index & $7.8\pm 0.5$ & $2.48^{+0.09}_{-0.05}$\\ \hline
\multicolumn{4}{l}{Gaussian Blurring ({\ttfamily gsmooth})}\\
$\sigma$ & keV & -- & $0.28^{+0.02}_{-0.04}$\\ \hline
\multicolumn{4}{l}{Cross-Normalization Constants (relative to FPMA)}\\
$C_{\rm XIS0}$ & -- & $1.081\pm 0.005$ & $1.081^{+0.002}_{-0.004}$\\
$C_{\rm XIS1}$ & -- & $1.038\pm 0.004$ & $1.038^{+0.002}_{-0.004}$\\
$C_{\rm FPMB}$ & -- & $1.001\pm 0.001$ & $1.001\pm 0.001$\\
$C_{\rm PIN}$ & -- & $1.205\pm 0.007$ & $1.205\pm 0.006$\\
$C_{\rm GSO}$ & -- & $1.17\pm 0.06$ & $1.16\pm 0.06$\\ \hline
$\chi^{2}/\nu$ & -- & 1512/1146 & 1510/1145\\ \hline
\end{tabular}
\end{center}
\end{minipage}
\end{table}

\end{document}